# MAD-FC: A Fold Change Visualization with Readability, Proportionality, and Symmetry

**Authors:** Bruce A. Corliss[1,2,*], Yaotian Wang[3], Francis P. Driscoll[1], Heman Shakeri[1], Philip E. Bourne[1,2]

**Affiliations:**

[1]School of Data Science, University of Virginia; Charlottesville, Virginia

[2]Department of Biomedical Engineering, University of Virginia; Charlottesville, Virginia

[3]Department of Statistics, University of Pittsburgh; Pittsburgh, Pennsylvania

*Corresponding author. Email: bac7wj@virginia.edu

**Short Title**: Mirrored Axis Distortion of Fold Change

**Classification**: bioinformatics, quantitative biology

**Keywords**: Bioinformatics, fold change, omics, data visualization, data science, genomics

## Abstract

We propose a fold change visualization that demonstrates a combination of properties from log and linear plots of fold change. A useful fold change visualization can exhibit: (1) readability, where fold change values are recoverable from datapoint position; (2) proportionality, where fold change values of the same direction are proportionally distant from the point of no change; (3) symmetry, where positive and negative fold changes are equidistant to the point of no change; and (4) high dynamic range, where datapoint values are discernable across orders of magnitude. A linear visualization has readability and partial proportionality but lacks high dynamic range and symmetry (because negative direction fold changes are bound between [0, 1] while positive are between [1, ∞]). Log plots of fold change have partial readability, high dynamic range, and symmetry, but lack proportionality because of the log transform. We outline a new transform and visualization, named mirrored axis distortion of fold change (MAD-FC), that extends a linear visualization of fold change data to exhibit readability, proportionality, and symmetry (but still has the limited dynamic range of linear plots). We illustrate the use of MAD-FC with biomedical data using various fold change charts. We argue that MAD-FC plots may be a more useful visualization than log or linear plots for applications that require a limited dynamic range (approximately ±2 orders of magnitude or ±8 units in log2 space).

## Introduction

Bioinformatics research often requires analyzing datasets that are expressed in units of fold change (*1*). This measurement type often represents the ratio of the sample mean of an experiment group divided by a control group. Fold change data is visualized to summarize the spread of the dataset and identify interesting datapoints- often those with the largest magnitude in either direction of change. Typically, scientists visualize fold change with a log plot or a linear plot (that directly presents raw fold change values) (*2*). Both plots have a unique set of properties that may be advantageous or disadvantageous depending on the particular purpose of the visualization.

We identify four visualization properties that are useful for interpreting fold change data. We define a new visualization, named mirrored axis distortion of fold change (MAD-FC), and illustrate how it has a combination of visualization properties from log plots and linear plots. We demonstrate the use of our visualization on real research data. Providing a novel fold change visualization may enhance the understanding and interpretation of fold change data in scientific research.

## Background

Charts are visual representations used to reveal patterns, trends, and relationships in datasets (*3*). They cater to a wide range of data analysis goals and datatypes, and as a result, exhibit a wide range of characteristics and properties. We define four properties that are specifically useful for visualizing fold change data: readability, proportionality, symmetry, and high dynamic range (Fig 1A). For defining these characteristics, we introduce the term *reference point*, which is the value in a fold change visualization that denotes no change and separates negative fold changes from positive fold changes. For a log plot, the reference point is zero, while it has a value of 1 for a linear plot. To aid with this discussion, we also define a linear encoding for fold change, *fold change from no change*, that simply converts fold change measurements to the number of fold changes from zero (Fig 1B). With this encoding, a fold change of 2 is 1 unit of fold change from no change, and a fold change of ½ is -1 units fold change from no change. We propose that the properties of a useful fold change visualization are:

1. **Readability**: a visualization with readability has a clear and direct mapping between the value of the datapoints and their spatial location. A linear visualization exhibits readability because the axis scale reflects the raw datapoint values, and the precise value of a datapoint is in proportion to its distance to the nearest axis labels (Fig 1C). A log plot would be considered partially readable by our definition- while the axis labels reflect the range of the dataset, the nonlinear spacing of datapoint values between axis tick marks from the log transform makes it more difficult to accurately discern datapoint values (Fig 1D). While datapoint values from a linear visualization is easily read from groups of points, care must be taken to decode each datapoint's value individually from a log plot.
2. **Proportionality**: a visualization exhibits proportionality if the fold change datapoints going the same direction are proportionally distant from the value denoting no change within the plot. This property allows for direct comparisons between the magnitude of positive fold changes separately from negative fold changes. A linear chart has proportionality for positive fold changes because the distance from the datapoint to the reference point is proportional to its value, but this relationship does not extend to negative fold changes (Fig 1C, light grey line). A log plot of fold change data does not exhibit proportionality because the distance from each datapoint to the reference point is a log transform of its value and not a linear mapping (Fig 1E, light grey line). This property makes it difficult to compare the magnitude of fold change datapoints based on their distance to the reference point.
3. **Symmetry**: a plot with symmetry has fold changes of the same magnitude equidistant to the point of no change regardless of their direction of change. This property allows scientists to visually compare the magnitude between positive and negative fold changes. A linear visualization does not have symmetry because negative fold changes are compressed between [0, 1] while positive are between [1, ∞] (Fig 1C, grey boxes to right of graph matching asymmetrically spaced positive and negative fold changes). A log plot exhibits symmetry because fold changes of opposite sign are equidistant to zero (i.e., a two-fold change and a half fold change are both equidistant to zero) (Fig 1E, symmetrically spaced grey boxes to right of graph).

4. **Dynamic Range**: fold change measurements can span across many orders of magnitude, and for some applications it is important to clearly discern the positions of points across this range. A log plot has high dynamic range because each unit on the axis scale represents an exponential increase of magnitude, allowing scientists to differentiate fold changes both close in value and across a large range of magnitudes (Fig 1D). A linear plot has much lower dynamic range, and typically can only differentiate datapoint values across a couple of orders of magnitude (Fig 1C).

We propose a fold change visualization, named *mirrored axis distortion of fold change*, that exhibits the readability of a linear plot (Fig 1G), proportionality (grey line overlapping with all fold change points), and symmetry (symmetrically spaced grey rectangles to the right of graph). Similar to a log plot, a transform is applied to the raw data and then the axis tick labels are then reassigned to ensure readability (Fig 1F).

## Methods

### MAD-FC Transform

We start with a fold change dataset with *n* measurements, $X_1, X_2,..., X_n$, and we wish to produce a visualization with readability, proportionality, and symmetry. We will perform two transforms on the raw data and then reverse the transforms on the axis tick labels to accomplish this. We start by defining symmetry between positive and negative fold changes. We accomplish this by defining the pairs of positive and negative fold change points that correspond to the same magnitude of fold change (Fig 2A). Reversing the direction of a fold change is simply the reciprocal of its value (i.e., $f(x) = 1/x$). To obtain symmetry, these pairs of points must become equidistant to some reference value.

Based on this table, we can define a transform that stretches the negative fold change values to match the spacing of the corresponding positive fold changes. We use the same transform that was used to reverse the direction of a fold change and then multiply by negative one. Since we only want to transform negative fold changes, we define a case equation that only transforms negative fold change measurements. We denote this mirror transform as $f_M$, where

$$f_M(x) = \begin{cases} x & if\ x \geq 1 \\ -\frac{1}{x} & otherwise \end{cases}. \qquad (1)$$

This transform shifts all negative fold changes to be symmetrically spaced from zero to their corresponding positive fold changes (MFC column in Fig 2A, visualized in Fig 2C). But this transform leaves a discontinuous region between [-1,1] where no fold change values can exist. This would give a misleadingly large spatial distance between negative and positive fold changes close to one (e.g., datapoints from rows 2-4 in Fig 2A) and for interval estimates that cross this region. We correct for this by translating all datapoints closer to zero by one unit, defined as a contraction transform $f_C$:

$$f_C(x) = \begin{cases} x - 1 & x \geq 1 \\ x + 1 & x < -1 \\ undefined & otherwise \end{cases}. \qquad (2)$$

This transform removes the discontinuous region, but the datapoint values no longer reflect the actual fold change values because they are shifted one unit (Con-MFC column in Fig 2A, visualized in Fig 2D). To correct for this, we will now reverse both transforms on the axis tick labels so that the datapoint's value can be read from the labels. We will perform this label reassignment by reversing the two transforms we performed on the data (i.e., $f_C(x)$ and $f_M(x)$). We denote axis tick labels as x' to clarify that the fold change data is not changed with these reversed transforms. We first reverse $f_C$ with the function $f^{-1}_C$:

$$f_C^{-1}(x') = \begin{cases} x' + 1 & x' \geq 0 \\ x' - 1 & x' < 0 \end{cases}. \qquad (3)$$

This transform changes the reference point of the graph from 0 back to 1-fold change (since the first case in $f^{-1}_C$ includes zero). We now reverse the mirrored fold change transform $f_M$ with $f^{-1}_M$, where

$$f_M^{-1}(x') = \begin{cases} x' & if\ x' \geq 0 \\ -\frac{1}{x'} & otherwise \end{cases}. \qquad (4)$$

After these transforms, the original FC values can be read from the chart (MAD-FC column in Fig 2A). We then can display the resulting negative fold change axis tick labels as decimals (Fig 2E), fractions (Fig

2F), or reciprocals (Fig 2G). We are left with a visualization that not only retains the readability of a linear visualization, but also exhibits proportionality and symmetry around a fold change of 1.

**Data and Material Availability**

Code used to generate all data and figures is written in R and available at:

https://github.com/bacorliss/mirrored_axis_distortion.

## Results

Different data visualizations can leave very different impressions of the same dataset and influence the conclusions drawn from the data. We demonstrate the importance of proper visualization with fold change measurements by comparing log plots, linear plots, and MAD plots of fold change with various biomedical datasets. We highlight the differences in their portrayal of the data, along with the unique advantages of using MAD-FC plots.

We start with an RNA-Seq dataset of differentially expressed genes from airway smooth muscle cells with and without glucocorticoid treatment to investigate mechanisms for their therapeutic effect for treating asthma (*4*). When performing differential gene expression analysis, it is common to visualize the data with a volcano plot and MA plot to identify interesting genes and summarize the trends. Volcano plots are a scatterplot that visualizes statistical significance versus effect size and enables a quick identification of genes that exhibit both high statistical significance and large fold change (*5*). We produce volcano plots with log, linear, and MAD plots of fold change for this dataset (Fig 3 A-C). While the log plot has symmetry and allows us to compare positive and negative fold changes, the magnitude of fold change datapoints are not proportional to their distance to the origin (Fig 3A). The linear plot allows us to compare the magnitude of positive fold changes, but all the negative fold changes are compressed between 0 and 1 (Fig 3B). The MAD plot of fold change allows for comparison between positive and negative fold changes symmetrically, along with comparing the magnitudes proportionally between fold changes of the same direction (Fig 3C). This type of plot is often used to prioritize candidate genes for further investigation. While the value of each individual datapoint can be read from a log plot, the spatial distribution of the point cloud gives a distorted summary when compared to their actual fold change values. The log fold plot gives a potentially misleading impression that there are many genes that are reasonably close to the max datapoint value for fold change, but the MAD plot highlights most of these candidates are less than 50% of the max fold change observed in the dataset. This is an important consideration when deciding which gene candidates should be prioritized for follow up studies.

MA plots are used to compare fold change of differential gene expression versus the average count of the same gene between both groups (essentially a Bland-Altman analysis for fold change data (*6*)). MA plots are used to highlight systematic bias or highly differentially expressed genes. For this case, the two groups are human airway smooth muscle cells cultured with and without glucocorticoid treatment (same dataset as Fig 3A-C). We produce MA plots with log, linear, and MAD fold change of this dataset (Fig 3 D-F). The advantages of the MAD-FC MA plot are the same as with volcano plots- the values of the datapoints are presented without the spatial distortion found with the log and linear plots.

The MAD-FC transform is especially useful for visualizing the sample distribution and uncertainty associated with fold change measurements (whether it's the standard deviation, standard error, quantile, distribution, confidence interval, credible interval, or support interval). To illustrate this, we produce a simulated dataset of fold change measurements with a 95% confidence interval that extends 2-fold change units above and below the point estimate for each study group. These study groups all have identical interval widths in fold change units, yet the MAD-FC plot is the only visualization where this consistency is apparent (Fig 4A-C). Both the log and linear fold change plots heavily distort the interval width, making it impossible to perceive that all the study groups have the same confidence interval widths. The MAD-FC visualization not only preserves the proportional distance of each datapoint to the reference point of no change, but also preserves the interval estimate width regardless of the value of the fold change measurement. This key behavior is apparent with a dataset measuring protein expression and phosphorylation in HER2-positive breast cancer cell lines treated with the MEK inhibitor refametinib (*7*) (Figure 4D-F). The linear visualization of fold change (Fig. 4E) used in the cited publication makes it easy to mistakenly conclude that the standard deviation appears approximately consistent across measurements and the negative fold change measurements are smaller in magnitude than the largest positive fold change measurement. However, when the same data is viewed by log plot (Fig 4D) or MAD plot (Fig 4F), the negative fold changes are revealed to have much larger standard deviations than the other measurements and they are comparable in magnitude to the largest positive fold change datapoint. The advantage with the MAD plot over the log plot is that the width of the bounds for standard deviation can be compared directly and proportionally between groups regardless of their distance from the reference point of no change.

The distortions exhibited by log and linear plots are even more pronounced in boxplots. A simulated datasets of boxplots with the sample median swept from 1/9 to 9-fold changes, with quartiles evenly spaced 2-fold changes between themselves, shows that not only the visualized width of the boxplots changes dramatically, but also the relative widths of quartiles within a single boxplot can become heavily distorted depending on the plot used (Fig 5A-C). This characteristic gives a false impression that the 4$^{th}$ and 6$^{th}$ study group have skewed distributions in fold change within the log plot (Fig 5A) when they are in fact symmetrical (Fig 5C). With boxplots of mRNA expression of various genes measured from patients with breast invasive carcinoma, ovarian serous cystadenocarcinoma, and lung squamous cell carcinoma, the appearance of the log, linear, and MAD fold change plots appear as if they represent entirely different datasets (datasets acquired from the Cancer Genome Atlas). Log plots of fold change exaggerate graphical features that are close to zero and compresses those that are further away. Linear plots of fold change heavily distort any features that extend into the region of negative fold change.

The same trends are observed when comparing violin plots between each of the three visualizations. With a simulated dataset of fold change distributions uniformly translated to different fold change values while maintaining the same degree of dispersion in fold change units (Fig 6A-C), only the MAD-FC plot reveals that each of these study groups have the same distribution shape, while the log and linear plots heavily distort the distribution depending on the fold change values. When displaying the same dataset as Fig 6D-F with violin plots, the overall appearance and trends of the datasets are again dramatically different for each of the visualizations (Fig 6D-F).

The MAD-FC visualization can also be useful for mapping fold change data to color gradients. A comparison of between heatmap of fold change for differential protein expression (Fig 7A-C) reveals that MAD-FC emphasizes measurements with the largest fold change values in the dataset. Visualizing fold change with a log mapping to color gradient makes it more difficult to clearly discern the largest fold change measurements (dataset from a study measuring Ubiquitin-protein interactors (*8*)).

## Discussion

Visualizations of fold change datasets are used to identify patterns, trends, outliers, and interesting datapoints in datasets. Different visualizations exhibit unique behaviors that highlight distinct aspects of the underlying data. Here we propose a visualization that enhances the usefulness of a linear visualization of fold change. A major shortcoming of linear fold change visualization is that negative fold change values are compressed from [0,1] and cannot be compared to positive fold change values by their spatial position. This limitation interferes with acquiring a holistic summary of fold change values.

Log plots allow the comparison of fold changes from both directions because positive and negative fold changes are positioned symmetrically about the origin. Yet the linear and proportional relationship between fold change value and spatial position is lost for log plots. MAD-FC plots are designed to maintain readability, symmetry, and proportionality. This combination of characteristics allows spatial position to be used as a proportional encoding for fold change value regardless of direction or magnitude of the measurement. While individual points can be read from a log plot, it is extremely difficult to identify trends from point clouds in a log scale. MAD-FC are designed to facilitate this process by preserving a linear and proportional encoding of fold change values. As a consequence, MAD-FC plots are especially useful for visualizing the distribution, quantiles, and interval estimates of fold change measurements since these visual features are not distorted in a spatially dependent manner as found with log and linear plots of fold change.

We enhance linear visualization of fold change with a transform strategy that is borrowed from log plots, where the datapoints and axis tick labels undergo different transformations. This produces a visualization that has a linear plot's ease of interpretation combined with the ability to compare the magnitude of fold change across all measurements. While this visualization lacks the high dynamic range found in log plots, in many applications comparing fold change across ±2 orders of magnitude (±8 units on the log 2 axis) is sufficient to identify interesting datapoints. This new visualization may be a useful tool for gaining a more intuitive summarization of datapoint values in fold change datasets. Different visualization yields unique summaries of the data that will influence which datapoints are prioritized for further investigation. A fold change visualization can also influence the value of the thresholds use to determine if a datapoint is considered noteworthy or not, which can significantly alter the interpretation of fold change data (*9*). Such visualizations could perhaps be used for other applications than what is shown here, such as meta-analysis techniques (*10*) and comparing effect size across broadly related experiments (*11–13*).

**Author Contributions:**

Conceptualization, Investigation, Writing- Original draft, Visualization, Data Curation, Software: BAC.
Methodology, Formal Analysis: BAC, FD.
Writing- Reviewing and Editing: BAC, FD, HS, YW, PEB.
Supervision: PEB.

**Acknowledgements:** This work was funded by PEB's endowment for the School of Data Science, University of Virginia.

**Declaration of Interests:**  The authors declare no competing interests.

# Figures

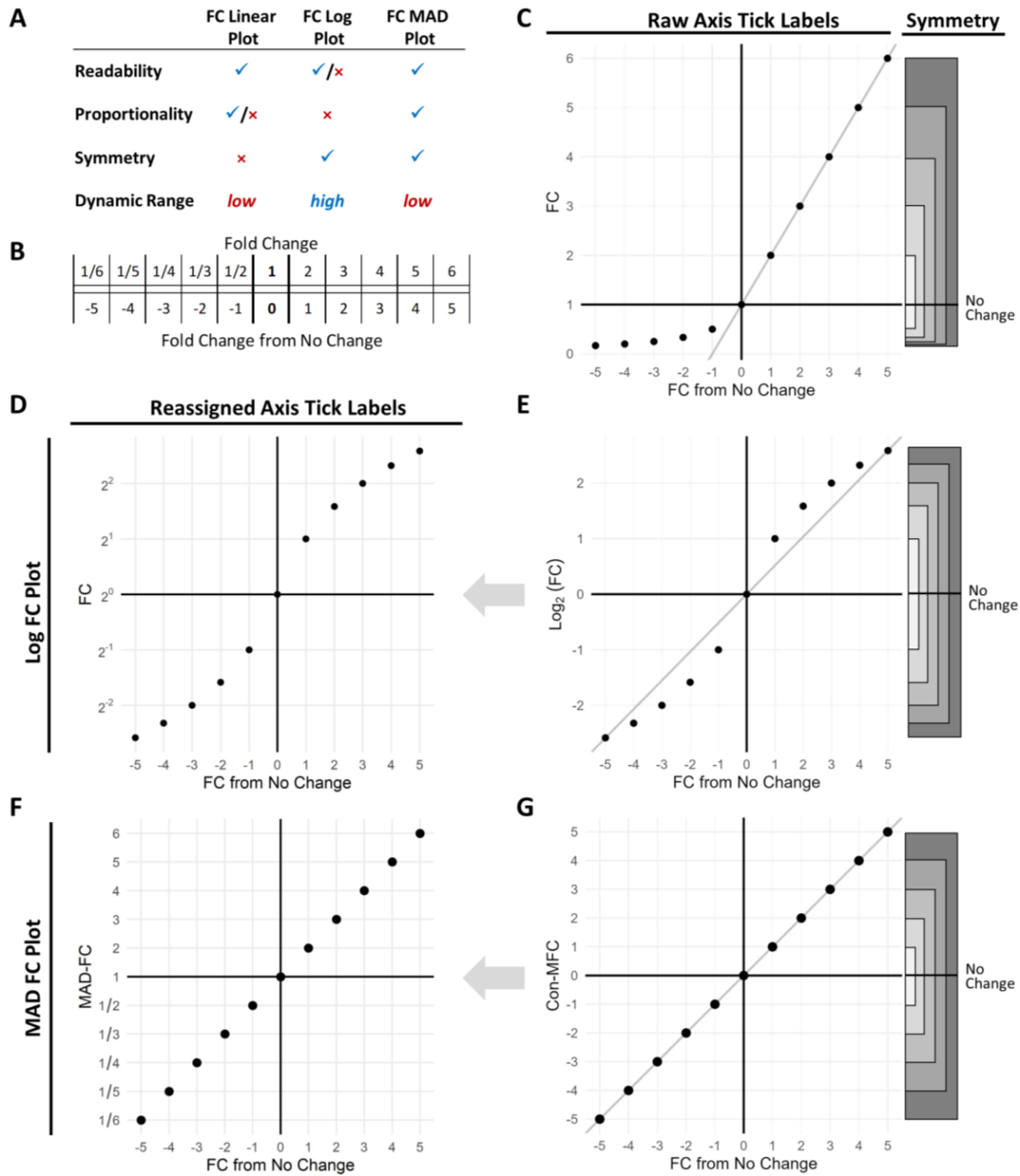

**Figure 1: Illustration of useful properties for visualizing fold change (FC) with a linear, log, and MAD plot**. (**A**) Table contrasting key properties between each of the plot types, and (**B**) illustration of the mapping between fold change values and fold change units from no change. (**C-G**) A fold change dataset is visualized with various charts to illustrate their characteristics, with FC values ranging from 1/6 to 6.

(**C**) A linear plot of FC lacks symmetry and is only proportional for positive FC values. (**D**) A standard log2 plot of FC data with y axis ticks relabeled to aid with reading the pre-transformed value for each datapoint. (**E**) A log2 plot with raw log2 transform values for fold change, showing symmetry about zero, but lacking proportionality due to the nonlinear transform. (**F**) MAD-FC plot with y axis ticks relabeled to aid with reading the pre-transformed value for each datapoint. (**G**) A MAD-FC plot with raw transform values, showing both symmetry about zero and proportionality. (**B**, **E**, **G**) grey line illustrates proportionality between each visualization and fold change units, and grey rectangles illustrate symmetry about reference point of no change by pairing corresponding positive and negative fold changes.

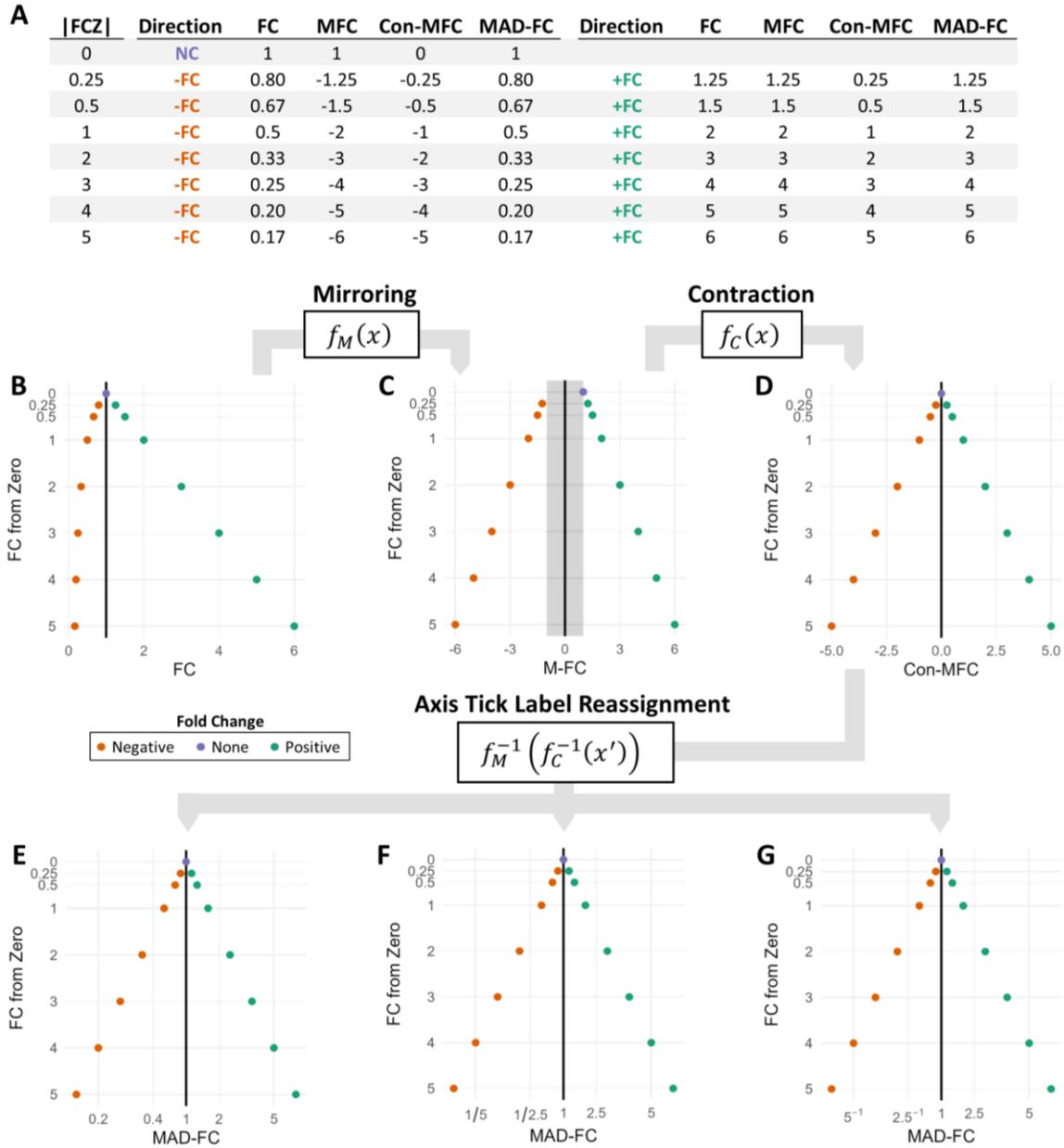

**Figure 2: Illustrations of transforms for MAD-FC Plot**. (**A**) Table of fold change datapoints that match negative fold changes (-FC) with their corresponding positive fold changes (+FC) (|FCZ| column denotes absolute value of fold change units from zero). (**B**) Plot of fold change datapoints in a linear scale, with negative fold changes compressed between [0 1] (datapoints from FC column). (**C**) Fold change values with a mirror transform applied ($f_M$) to the negative fold changes to stretch their position to match the corresponding positive fold changes (grey rectangle denotes undefined region between (-1 1), datapoints from MFC column). (**D**) A contraction transform ($f_C$) pulls both positive and negative fold changes 1 unit

closer to zero, eliminating the undefined region, but leaving fold change labels shifted 1 unit from their actual value (datapoints from Con-MFC column). (**D**) The transforms in (C) and (D) are reversed on the axis tick labels to annotate the datapoints with their actual fold change value. These steps in aggregation represent a Mirrored Axis Transform of Fold Change (MAD-FC) and can be annotated as a (**E**) decimal, (**F**) fraction, or (**G**) exponent. MAD-FC datapoints have the same values as the original FC points, but they are spatially distorted to achieve symmetry and proportionality.

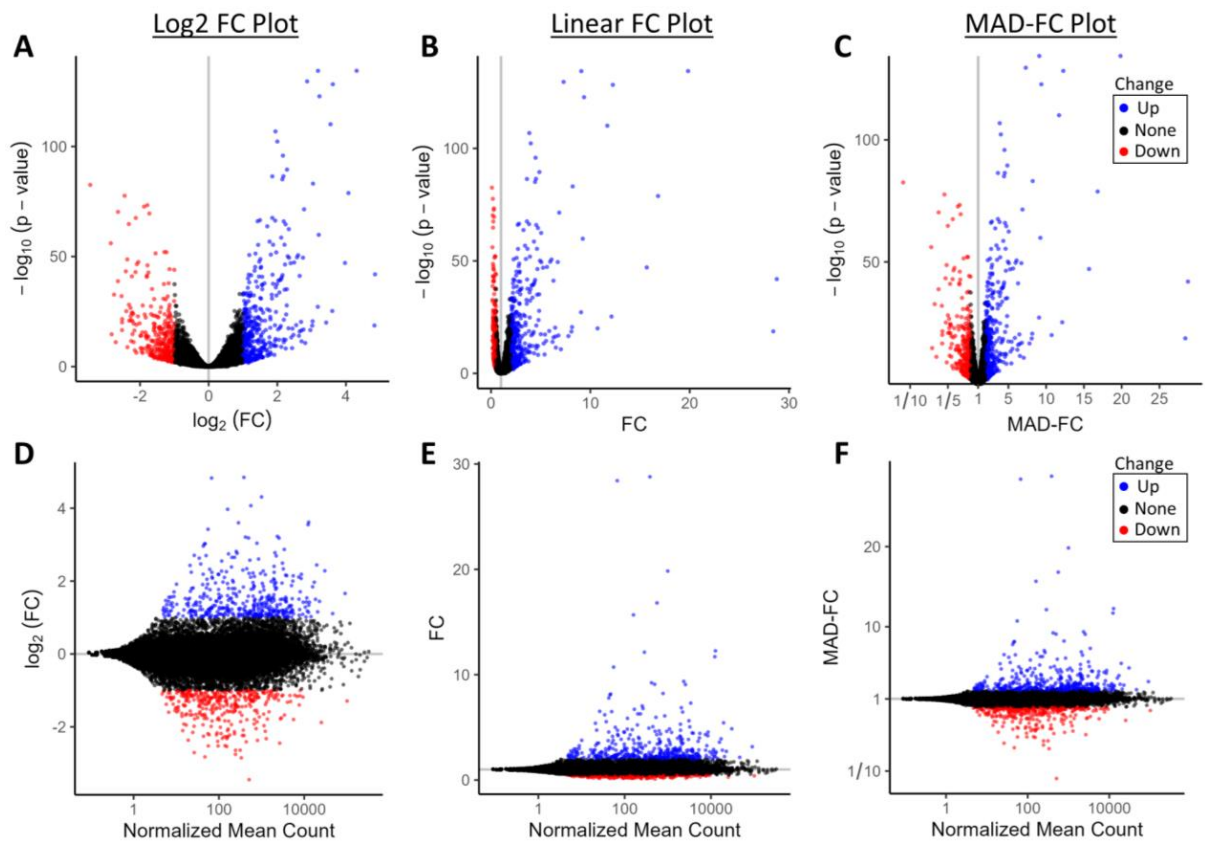

**Figure 3: Comparison of log, linear, and MAD fold change plots for RNA-Seq data**. Volcano plots of p-value versus (**A**) log, (**B**) linear, and (**C**) MAD fold change. MA plots using (**E**) log2, (**F**) linear, and (**G**) MAD fold change versus normalized mean count.

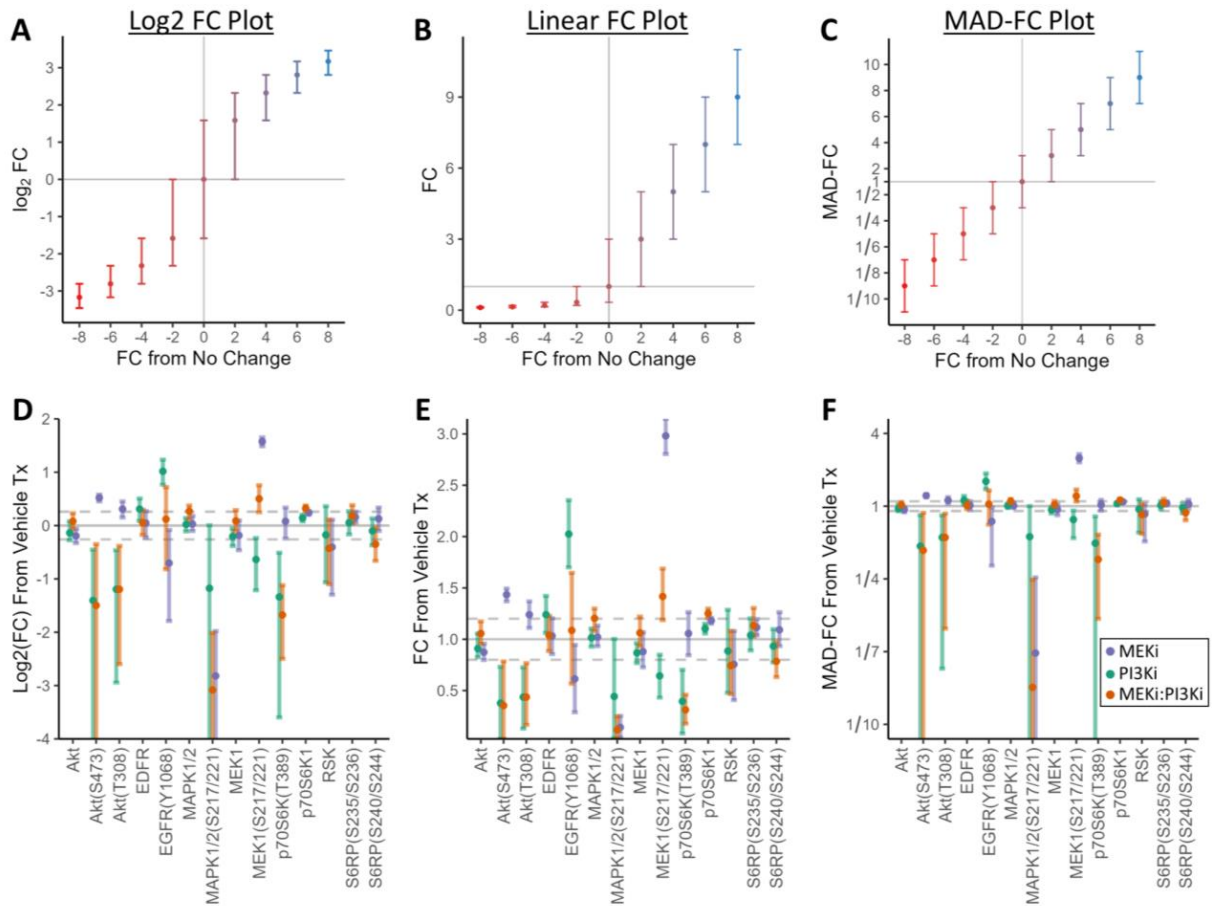

**Figure 4: Comparison between fold change plots with interval estimates**. Fold change interval estimates with the same interval width across groups visualized with a (**A**) log2, (**B**) linear, and (**C**) MAD fold change plot (synthetic dataset with interval estimates spanning from -2 to +2-fold change units from the point estimate, error bars are confidence intervals). Comparison of (**D**) log, (**F**) linear, and (**F**) MAD fold change plot of protein expression and phosphorylation in HCC1954-P cells treated with either 300nM refametinib (MEKi) or 15nM copanlisib (PI3Ki) alone or in combination (MEKi - 300nM: PI3Ki - 15nM) (error bars are standard deviation).

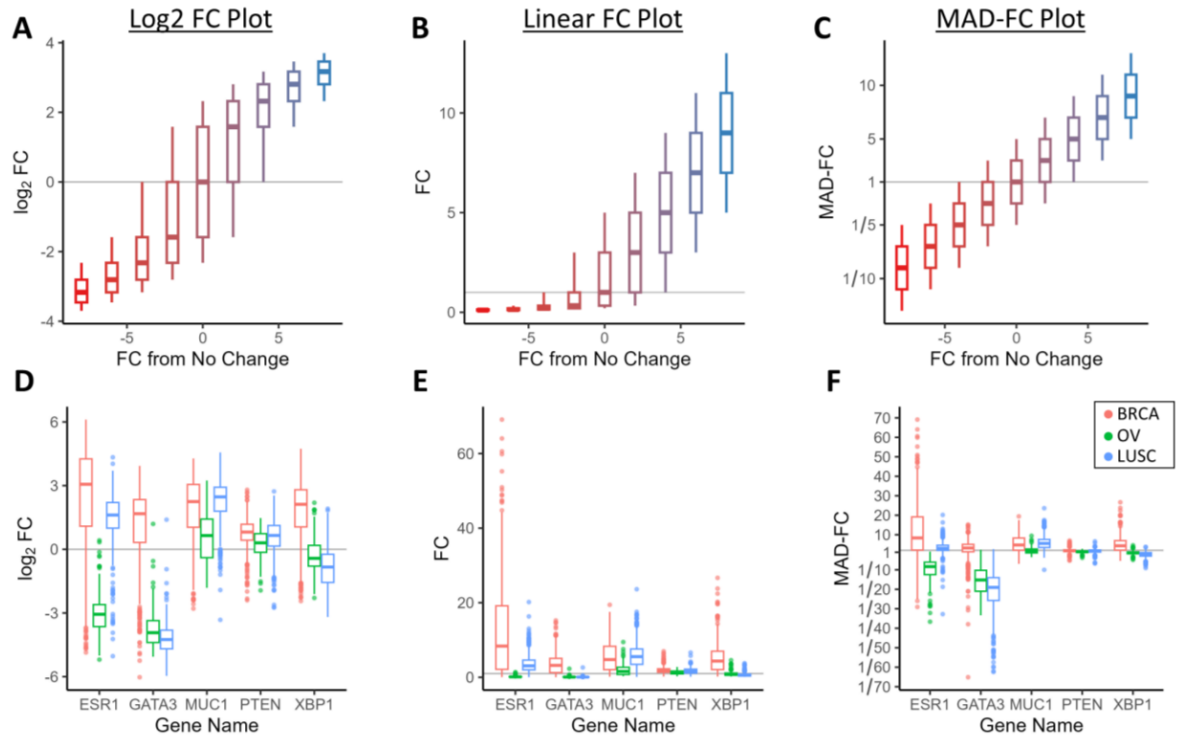

**Figure 5: Comparison between fold change box plots**. Fold change boxplots of simulated data with the same quantile widths across groups, visualized with a (**A**) log, (**B**) linear, and (**C**) MAD fold change plot (2-fold change unit differences between each quantile boundary). Comparison of (**D**) log, (**F**) linear, and (**F**) MAD fold change plot of mRNA expression of various genes measured from patients with breast invasive carcinoma (BRCA), ovarian serous cystadenocarcinoma (OV) and lung squamous cell carcinoma (LUSC) (datasets from the Cancer Genome Atlas).

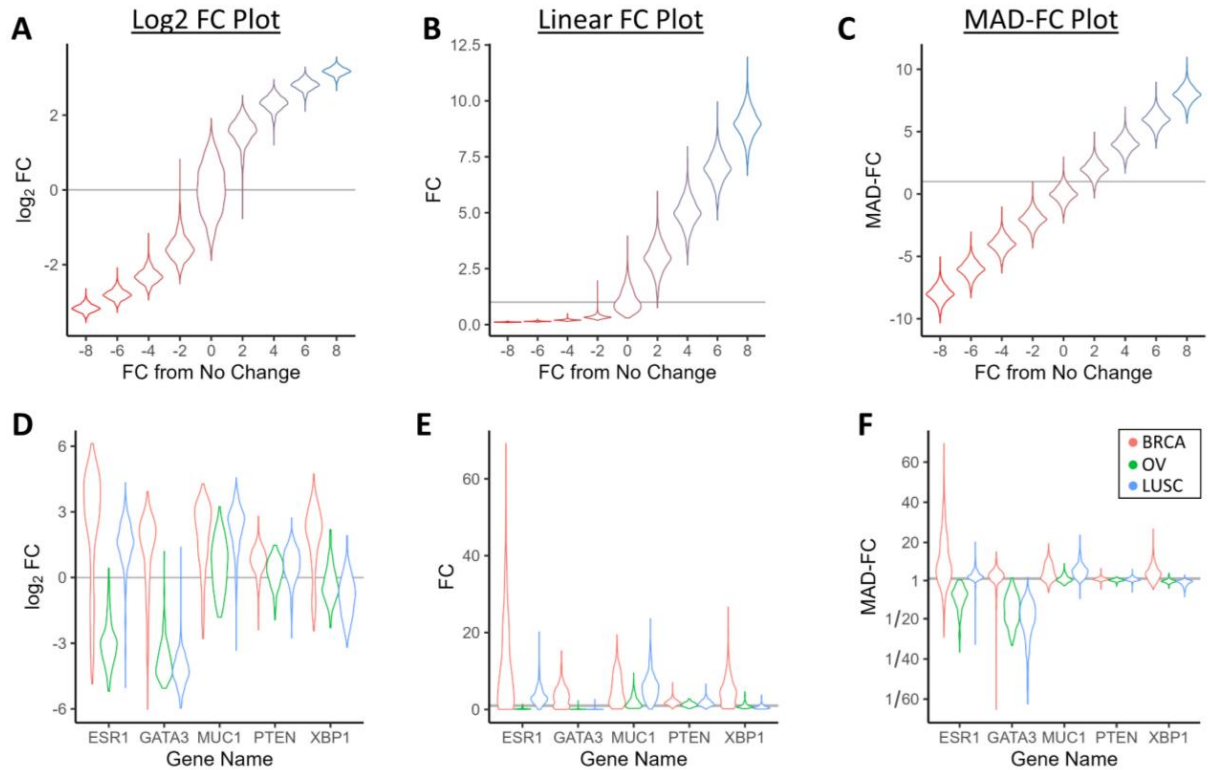

**Figure 6: Comparison between fold change violin plots**. Fold change violin plots with the same dispersion of measurements across groups visualized with a (**A**) log, (**B**) linear, and (**C**) MAD fold change plot (simulated dataset with identical dispersion in fold change units across all groups). Comparison of violin plots with (**D**) log, (**F**) linear, and (**F**) MAD fold change of mRNA expression of various genes measured from patients with breast invasive carcinoma (BRCA), ovarian serous cystadenocarcinoma (OV), and lung squamous cell carcinoma (LUSC) (datasets from the Cancer Genome Atlas).

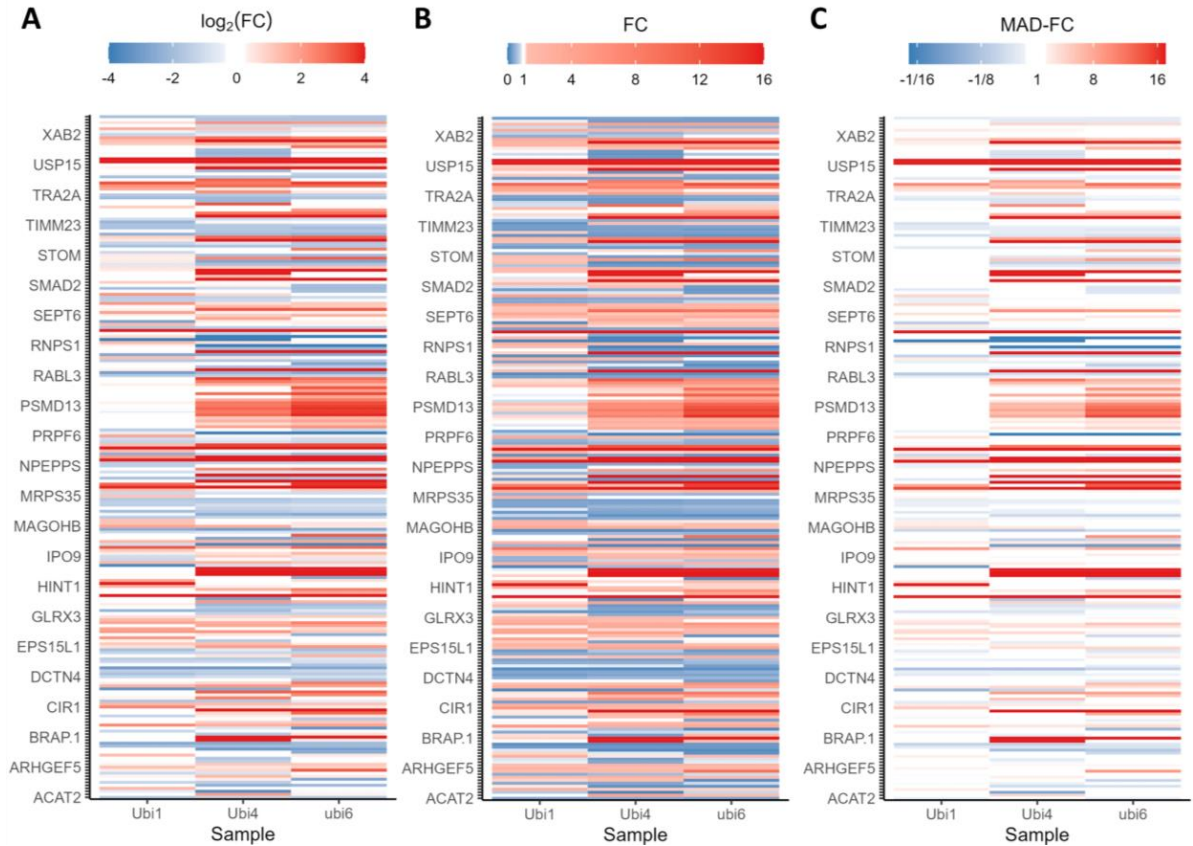

**Figure 7: Comparison of heatmaps with different encodings between fold change and color.** Comparison of heatmaps with (**A**) log, (**B**) linear, and (**C**) MAD fold change color mapping of protein expression from a dataset measuring several Ubiquitin-protein interactors.